\documentclass[usenatbib]{mn2e}

\usepackage{epsfig}
\usepackage{threeparttable}
\newcommand\apj{ApJ}%
\newcommand\apss{Ap\&SS}%
\newcommand\aap{A\&A}%
\newcommand\mnras{MNRAS}%
\newcommand\pasp{PASP}%
\newcommand\pasj{PASJ}%
\newcommand\araa{ARA\&A}%
\begin{document}
\title{The 2009 outburst from the new X-ray transient and black-hole candidate XTE J1652$-$453}
\author[Peng Han et al.]
       {Peng Han$^{1}$\thanks{E-mail: hanp@mail.ihep.ac.cn}, Jinlu Qu$^{1}$, Shu Zhang$^{1}$, Jianmin Wang$^{1,2}$, Liming Song$^{1}$,
       \newauthor Guoqiang Ding$^{3}$, Shuping Yan$^{3,1}$, and Yu Lu$^{1}$\\
       $^{1}$Laboratory for Particle Astrophysics, Institute of High Energy Physics, Chinese Academy of Sciences, 19B Yuquan Road,\\ Beijing, 100049, China\\
       $^{2}$Theoretical Physics Center for Science Facilities (TPCSF), CAS\\
       $^{3}$Urumqi Observation, National Observatories, Chinese Academy of Sciences, South Beijin Road, Urumqi, Xinjiang 830011, China}
\date{}
\pubyear{2010}
\pagerange{\pageref{firstpage}--\pageref{lastpage}}

\maketitle
\label{firstpage}

\begin{abstract}
The  RXTE and Swift observations on the 2009 outburst from a newly discovered transient and black-hole candidate XTE J1652$-$453 are analyzed. The source was observed by RXTE to behave a sequence of spectral states that are typical to the black hole XRBs.  During the first 7 observations, the source is diagnosed to stay in a high/soft state,  from  the spectrum dominated  by soft thermal component which contributes an average of $\sim85\%$ to the X-ray flux at  2--20keV, and from the hardness $\sim$ 0.1 showing up in the hardness-intensity diagram (HID). For the last 20 observations the spectral state is classified as {\sl low/hard} according to an average hardness of $\sim$ 0.8 and a balance between the thermal and the non-thermal components:  a power-law component takes $\geq80\%$ of total 2--20keV flux. Located in between is an {\sl intermediate} state that the source might have experienced. The usual relationship between rms and hardness presents in HRD as well.  Throughout the outburst no QPOs are found in XTE J1652$-$453.
\end{abstract}


\section{Introduction}\label{sec:intro}
The XTE J1652$-$453 was discovered as a new transient source on 2009 June 28 and July 01 in two PCA monitoring scans of the galactic ridge region with the Rossi X-ray Timing Explorer (RXTE) observatory \citep{ATel2107}. The two observations showed a quickly rising flux, and the energy spectrum suggested a black hole candidate in a spectral ``{\sl high}'' state. Then the Swift observed XTE J1652$-$453 in Photon Counting mode on 2009 July 03, and the spectrum can be fit with an absorbed blackbody model, with a column density of $(4.1$--$5.5)\times10^{22}$cm$^{-2}$ and a color temperature of KT$=$(0.5--0.7)keV. Based on the improved analysis of the Swift XRT image, the X-ray position was fixed to R.A.$=$16h52m20.33s, Dec.$=$ $-$$45\,^{\circ}$20'39.6'', with an estimated $90\%$ error radius of 2.5 arcsec \citep{ATel2108,ATel2120}. In addition to the X-ray observations, there were at least two objects(named star A and star B) visible in the K's bands image within the Swift error circle from the observations with the Infrared Survey Facility 1.4m telescope at SAAO\citep{ATel2125}. However, in the follow-up near-infrared observations, the star B showed no significant variabilities, suggesting that it could be an unrelated interloper star\citep{ATel2190}. Around 2009 September 10, the BAT monitor aboard Swift showed an increase of hard X-ray count rate from XTE J1652$-$453 within a couple of days, that may be indicative of a transition back to the hard state. The spectral evolution observed from 2009 September 14 to September 20 was consistent with XTE J1652$-$453 going through a transition to the hard state. The spectrum observed with Swift on 2009 September 23 can be described by a simple absorbed power law with an hydrogen column density of $(4.9\pm0.8)\times10^{22}$cm$^{-2}$ and a photon index of $1.97\pm0.3$, which is an attribute of spectrum of hard state\citep{ATel2219}.

From the view of observation, black hole candidates (BHCs) can be divided into at least two distinct states: the {\sl low/hard} state (LHS) and the {\sl high/soft} state (HSS) based on their energy spectra behavior during the outbursts. There are two main intermediate sates between the LHS and HSS: the {\sl hard intermediate} state (HIMS) and {\sl soft intermediate} state (SIMS), distinguished in the hardness-intensity diagrams (HIDs). An additional X-ray state named {\sl very high} state was identified as the spectra comprised of both a thermal component and a power law component which was steeper than the LHS\citep{Miyamoto1993, Zhang1997}. Both of the spectral and temporal properties of these sources vary from one state to another state. Many models are developed to understand the nature of BHCs depend on the spectral properties\citep[for a review, see][]{ McClintock2004, Homan05, Remillard06, Belloni09}. All these models consider the spectra as generated by two main components: thermal and nonthermal. The HSS corresponds to the soft spectrum, dominated by a thermal component. The LHS is dominated by a nonthermal emission, basically associated only to the early and late phase of an outburst. Therefore, it is important to follow the long time spectral evolution of individual sources to have deep research on their state transitions and properties.

Definitely, there are states transitions during the evolution of XTE J1652$-$453 from the observations with Swift and RXTE\citep{ATel2107, ATel2120, ATel2219}. The spectra from the outburst to decay can be got from the RXTE observations. Therefore, it is very convenient for us to have a comprehensive and systematic study on the state transitions also with spectral and temporal properties of this new black hole candidate.

In this paper we present the RXTE and Swift data analysis of the new X-ray transient XTE J1652$-$453. In Section 2 we describe the RXTE and Swift observations and data reduction. The energy spectral and timing analysis are reported in Section 3, then the evolution of the spectral states are studied. Our conclusions and implications are discussed in Section 4.

\section{Observation and data reduction}
All the publicly available data obtained from RXTE on XTE J1652$-$453 were used in this work. These data covered the period from July 2009 when this source was first observed to November 2009 including 55 observations, with the identifier (OBSID) of proposal number P94044 and P94432 in the data achieve of High Energy Astrophysics Science Archive Research Center (HEASARC) (see Table 3). These observations sum up $\sim10$ks of exposure time on this source with each of them lasting from $\sim1$ to several ks.

For the RXTE data reduction we used the standard \textsc{FTOOLS} package. Unfortunately, the background of some observations with HEXTE can not be subtracted. Therefore the 16--s time-resolution Standard 2 data from the Proportional Counter Array (PCA) were used to do the following studies. To precise, only the PCU2 data were used for the spectral analysis, since only the PCU2 was totally on during the observations. We filtered the data using the standard RXTE/PCA criteria. The time intervals were picked up under the constraints on elevation angle $>10\,^{\circ}$, and pointing offset $<0.02\,^{\circ}$. For the PCA background estimations, we used the most recent available background model from the HEASARC website for bright sources\footnote{pca\_bkgd\_cmbrightvle\_eMv20051128.mdl}. During the energy spectra analysis, only the data of 8--63 PCA absolute energy channel were used because of their lower uncertainties.

In addition to the RXTE observations, there are 4 observations with Swift/XRT shown in Table 1 with three of these are in Photon Counting (PC) mode and the other one in Windowed Timing (WT) mode \citep[the 2009 July 03 and September 23 observations see][]{ATel2107, ATel2120, ATel2190}. The Swift/XRT data were processed with the standard xrtpipeline tool and the level 2 event files were standardly filtered. For the heavily pile-up on July 03 observation\citep{ATel2120}, a 30 arcsecond exclusion radiu was used in extracting the spectrum.

\setcounter{table}{0}
\begin{table*}
\begin{minipage}{155mm}
\begin{center}{
\scriptsize
\caption{Journal of Swift/XRT observations of XTE J1652$-$453}\label{tab:line-models}
\begin{tabular}{cccccc}

\hline\hline
\multicolumn{1}{c}{Obs. ID} &
\multicolumn{1}{c}{Date, UT} &
\multicolumn{1}{c}{Time start} &
\multicolumn{1}{c}{MJD} &
\multicolumn{1}{c}{XRT exp.(s)} &
\multicolumn{1}{c}{Obs. Mode}  \\
  &   &   &   &  (s) &  \\
\hline
00031440001  & 2009-07-03  & 13:20  & 55015.56  &1361.4  &\textsc{PC} \\
00031440002  & 2009-07-03  & 14:42  & 55015.61  &3812.3  &\textsc{WT} \\
00031440003  & 2009-09-23  & 14:38  & 55097.61  &1988.5  &\textsc{PC} \\
00031440004  & 2009-10-03  & 04:27  & 55107.19  &1981.6  &\textsc{PC} \\
\hline
\end{tabular}
\normalsize}
\end{center}
\end{minipage}
\end{table*}

All the spectra are fitted with XSPEC v12.5.1 and the model parameters are estimated with 90\% confidence level.

\section{results}
\subsection{Intensity and hardness}

\setcounter{figure}{0}
\begin{figure}
\begin{center}
\hspace{-0.68cm}
\resizebox{1.08\columnwidth}{!}{\rotatebox{0}{\includegraphics[clip]{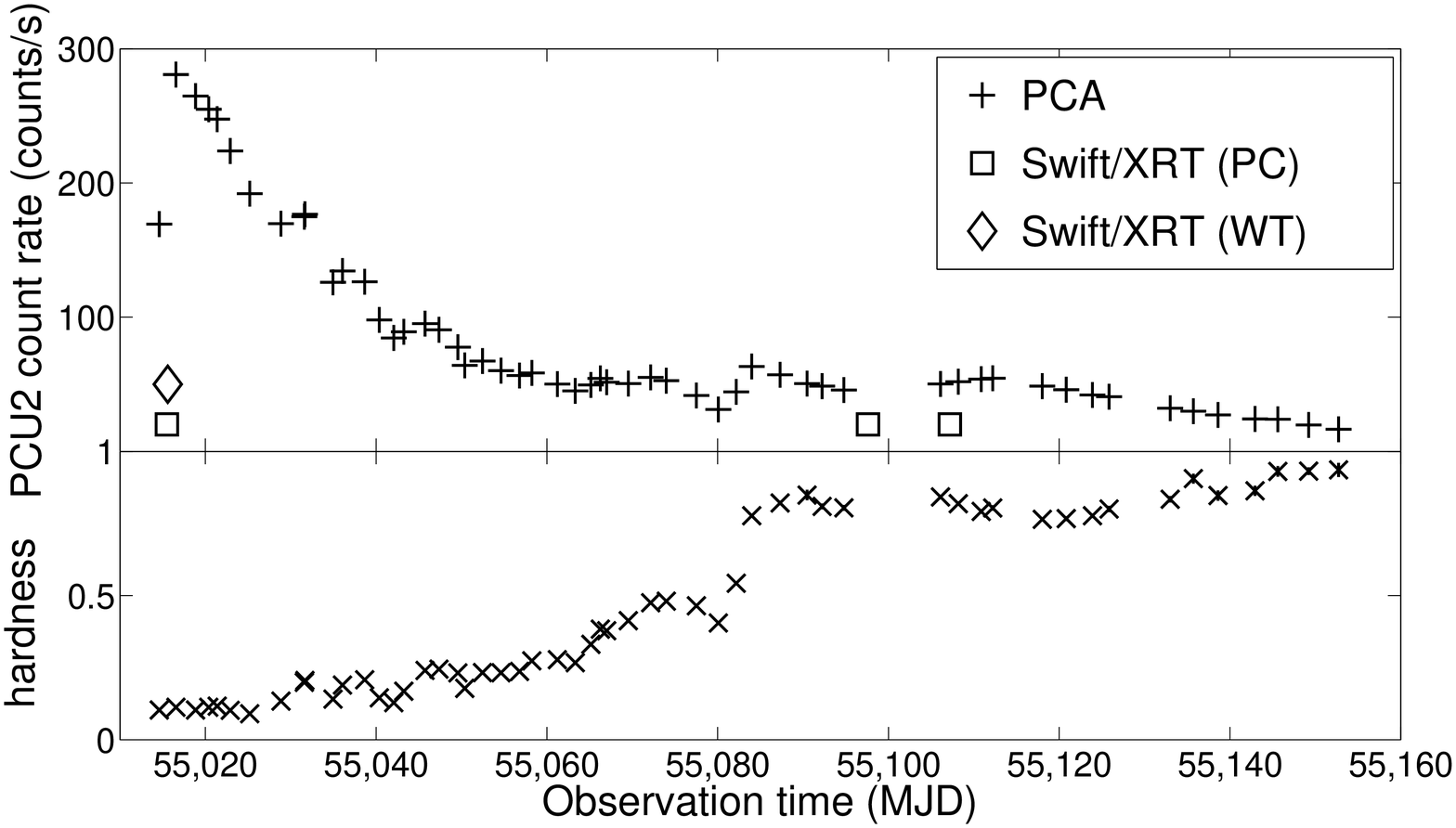}}}
\caption{The PCA lightcurve and hardness ratio (6.12--10.22keV$/$3.28--6.12keV) evolution of XTE J1652$-$453. Only the observation time of Swift/xrt
used in this work are marked with squares (PC mode) and diamond (WT mode) without any information of the Swift count rates.}
\label{fig:lightcurve}
\end{center}
\end{figure}

The PCA lightcurve and hardness of XTE J1652$-$453 during the observations is shown in Figure \ref{fig:lightcurve}. Each observation is represented  with a single symbol. To be consistent with the energy spectra analysis, the intensity is defined as the PCU2 count rate of Standard 2 data in the 4 -- 54 energy channel (about 3.28--26.55keV energy band). The hardness is defined as the count rate in the 11--20 Standard 2 channel (6.12--10.22keV energy band) divided by the count rate in the 4--10 Standard 2 channel (3.28--6.12keV energy band). According to the RXTE/PCA data, the XTE J1652$-$453 shows a classic light curve with a quick rise of flux and a comparatively slower decay\citep{Chen97, Wu2010}. It looks very like the 2002 outburst of 4U 1543$-$47\citep{Kalemci05, Reig06}.

From the beginning of the outburst, the hardness remained the same in the first several observations until MJD $\sim55025$, and at that point it started to increase slowly, roughly linearly with time. Around MJD $\sim55080$, a dip appeared in the hardness diagram with a dip in the lightcurve in the same time. After that, the hardness increased rapidly by a factor $\sim2$ in a few days, then it kept in a high level with an approximately constant value.

The hardness-intensity-diagram (HID) of the outburst of XTE J1652$-$453 is shown in top panel of Figure \ref{fig:hid}. Since only the decay part of the outburst was observed by RXTE, the whole evolution track is present as the end part of the {\sl q} like diagram compared to GX339$-$4 \citep{Belloni05, Belloni09, Hiemstra2010}. It looks very like the 2008 January outburst of BHC H1743$-$322 \citep{Capitanio09}. At the beginning of the outburst, the intensity of XTE J1652$-$453 was high, and the spectrum was very soft (for detail, see 3.2). First several observations were located in the upper left part of the HID. With the overall decrease of intensity and increase of hardness, XTE J1652$-$453 followed the track from upper left to lower right in the HID. Then it stayed in the right branch during the last several observations at the end of the decay.

\setcounter{figure}{1}
\begin{figure}
\begin{center}
\hspace{-0.68cm}
\resizebox{1.07\columnwidth}{!}{\rotatebox{0}{\includegraphics[clip]{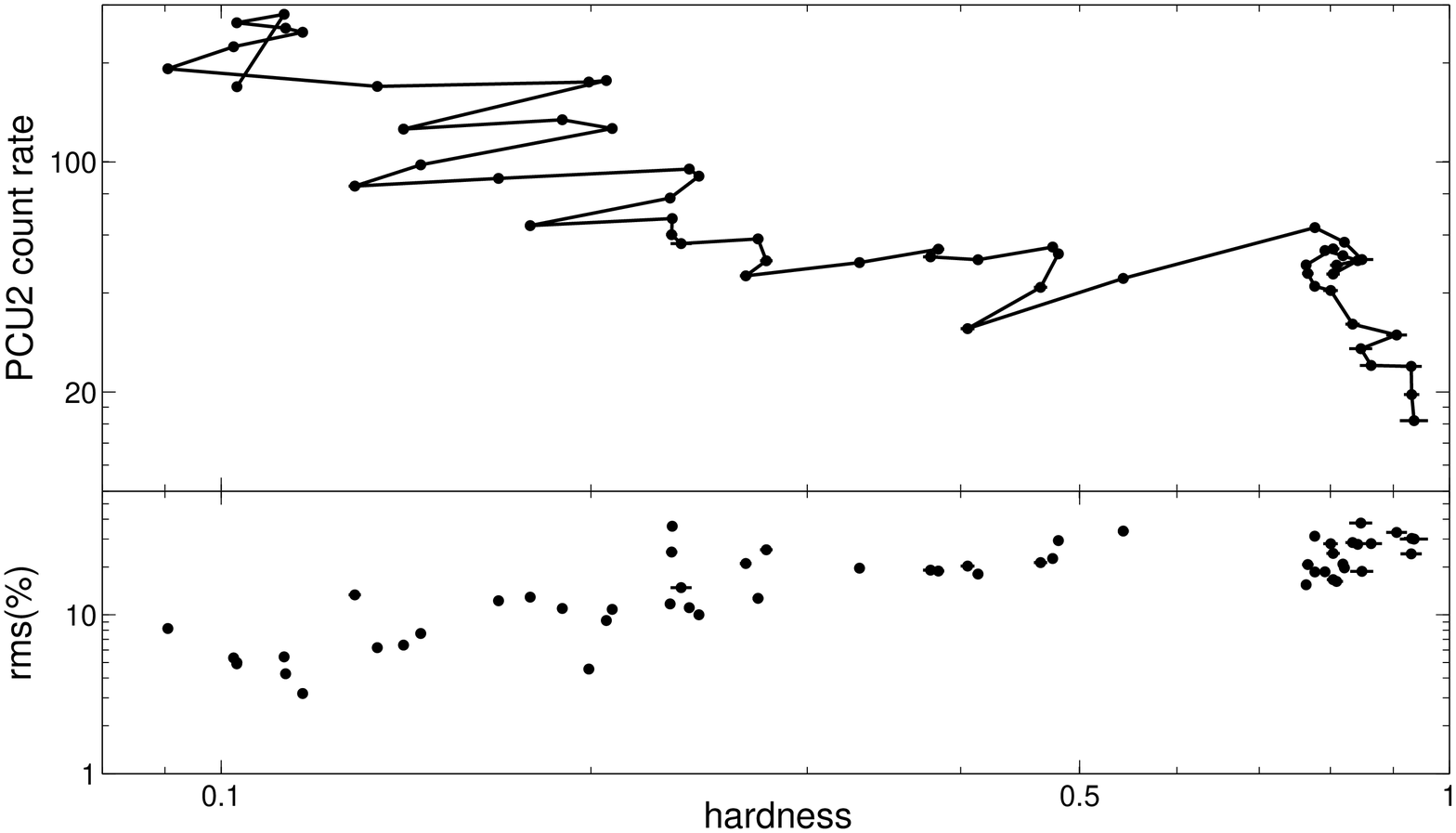}}}
\caption{Top panel: HID of all the RXTE/PCA observation of XTE J1652$-$453 during the
2009 outburst. Bottom panel: HRD of the same observations.}
\label{fig:hid}
\end{center}
\end{figure}

\subsection{Energy spectra analysis}
Many models have been developed to describe the spectral states of BHBs, and it has been known that the energy spectra of BHBs often exhibit a composite shape comprising of a thermal and a nonthermal components\citep[for a review, see][]{McClintock2004, Remillard06}. The thermal component is well modeled by a multitemperature blackbody, which originates in the inner accretion disk\citep{Shakura1973, Novikov1973, Mitsuda84}. The nonthermal component is usually described by a power law, which generally extends to much higher energies than the thermal component. Sometimes the spectra suffer a break or some kind of cutoff at high energies if the spectra are of high signal-to-noise out to high enough energies, and these spectra are generally produced with thermal Comptonization models\citep{Thorne1975, Shapiro1976, Narayan1995}. However, the thermal Comptonization process can be well modeled by a power law over a wide range of energies between the seed photon energy and the cutoff energy.  Although the simple model has significant limits, it is applicable and effective in characterizing the spectra of a single source or BHBs. During the state transitions of BHBs, one or the other component may dominate the X-ray luminosity\citep{Belloni09}. It is very important to follow the long time spectral evolution of individual sources, and construct an unified model to study the spectral properties during the observations.

The hydrogen column density $N_{H}$ is a very important parameter in X-ray energy spectra fittings. The absorption is non-negligible especially in the energy band below $\sim 10$ keV. The hardness and soft X-ray flux may be affected significantly by different $N_{H}$ in each spectral approximation. Because of the limited spectral resolution of RXTE, we used the Swift/XRT observations to confirm the hydrogen column density. Since the spectrum of 2009 July 03 observation was consistent with a stellar mass black hole in the canonical {\sl high/soft} state \citep{ATel2120}, we used the model of a multitemperature disc blackbody with the interstellar absorption to fit the spectrum. The Swift observation on 2009 September 23 was believed to be in the hard state after transition\citep{ATel2219}, the model of absorbed power law can give a good fit to the spectrum together with the October 03 observation. The best fit parameters for the observations is presented in Table 2.

\setcounter{table}{1}
\begin{table*}
\begin{minipage}{155mm}
\begin{center}{
\scriptsize
\caption{Summary of spectral fit parameters for the Swift/XRT PC mode observations.}\label{tab:line-models}
\begin{tabular}{cccccc}
\hline\hline
Observation ID    &$N_{H}$    &$T_{d}$    &$\Gamma_{pl}$    &reduced $\chi^{2}$    &d.o.f \\
                  &$10^{22}cm^{-2}$    &keV    &photon index    &        &\\
\hline
00031440001  &$4.0^{+0.14}_{-0.13}$  &$0.80\pm0.02$  &      &1.2    &294\\
00031440002  &$3.8^{+0.09}_{-0.09}$  &$0.80\pm0.01$  &      &1.4    &372\\
00031440003  &$4.2^{+0.47}_{-0.42}$  &               &$1.8\pm0.14$  &0.98  &158\\
00031440004  &$3.9^{+0.40}_{-0.38}$  &               &$1.7\pm0.13$  &1.1   &176\\
\hline
\end{tabular}
\normalsize}
\end{center}
\end{minipage}
\end{table*}

The values of $N_{H}$ in our fitting are comparable with the results of Markwardt and Coriat et al\citep{ATel2120, ATel2219}. Furthermore, the $N_{H}$ can not be constrained by RXTE/PCA data. Therefore, the hydrogen column density $N_{H}$ was fixed to be the same during all the spectral fits of total 55 RXTE observations with a value of $4.0\times10^{22}$cm$^{-2}$ based on the Swift spectra analysis.

For the spectra fitting of total 55 RXTE observations of XTE J1652$-$453, we basically use a model consisting of a power law component, a multitemperature disc blackbody. A systematic error of $0.5\%$ was added to the data, and most of the spectra are well described by this basic model. Figure 3 shows an example of the RXTE observation (obsID 94432--01--04--00, MJD $\sim$ 55065) which lasted for $\sim 14$ksec. Based on the previous studies of the black hole binaries, their X-ray energy spectra sometimes show broad Fe emission lines at around $\sim6-7$keV\citep[for a review, see][]{Miller07}. Besides, a broad Fe emission line was observed with XMM-Newton during the evolution of XTE J1652$-$453\citep{Hiemstra2010}. To give a best fit to the PCA spectra of XTE J1652$-$453, an additional gaussian shape emission line at 6.4keV was added into the basic model. However, this line feature was typically fitted to have zero flux during the spectral analysis of total 55 observations. Therefore, we believe that the basic two component model is feasible in our spectral analysis, and the spectral properties can be contrasted reasonably by using this unified model. Then the fitting results are used to study the correlation of spectral and variability parameters. The best fit parameters for the RXTE observations can be seen in Table \ref{tab:line-models}.

The evolution of some best fit spectral parameters with observation time are present in Figure \ref{fig:spectra}. Based on our spectral analysis, both the power law photon index and the temperature of disc blackbody have a overall decrease with observation time and the count rate. Around MJD $\sim55080$, the soft component fraction has a sharp jump with a slightly increase of the power law photon index, then they decrease significantly in a few days.

\setcounter{figure}{2}
\begin{figure*}
\begin{center}
\hspace{-0.3cm}
\resizebox{1.025\columnwidth}{!}{\rotatebox{-90}{\includegraphics[clip]{spec.ps}}}
\hspace{0.4cm}
\resizebox{0.98\columnwidth}{!}{\rotatebox{-90}{\includegraphics[clip]{model.ps}}}

\caption{\textit{Left:} RXTE/PCA spectrum of XTE J1652$-$453 (obsID 94432--01--04--00), in the bottom panel, the best fitting residual
of the model consisting of a power law and a multitemperature disc blackbody.\textit{Right:} The model used in the spectral fitting.}
\label{fig:example}
\end{center}
\end{figure*}

\setcounter{figure}{3}
\begin{figure}
\begin{center}
\resizebox{1.07\columnwidth}{!}{\rotatebox{0}{\includegraphics[clip]{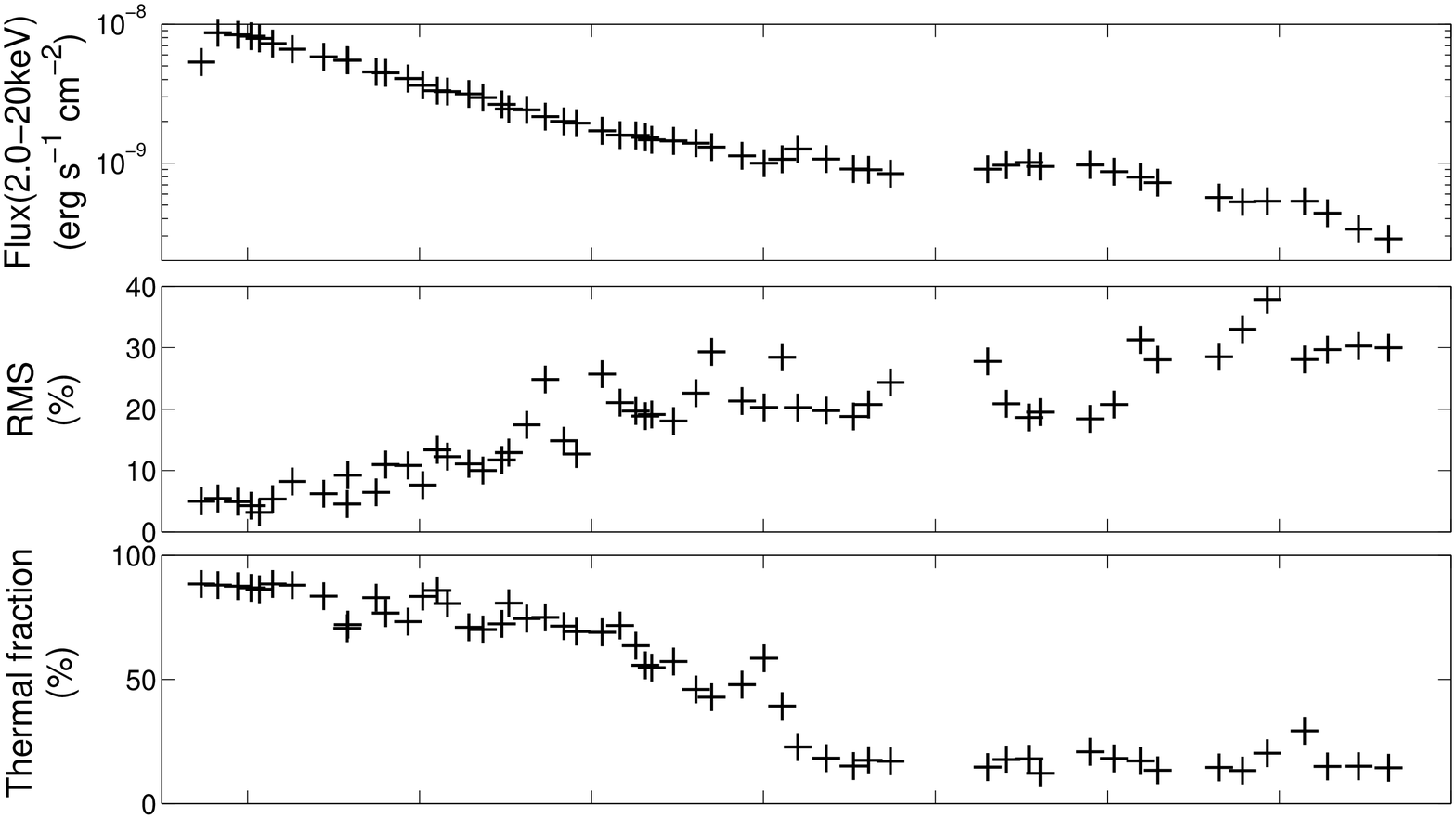}}}
\resizebox{1.07\columnwidth}{!}{\rotatebox{0}{\includegraphics[clip]{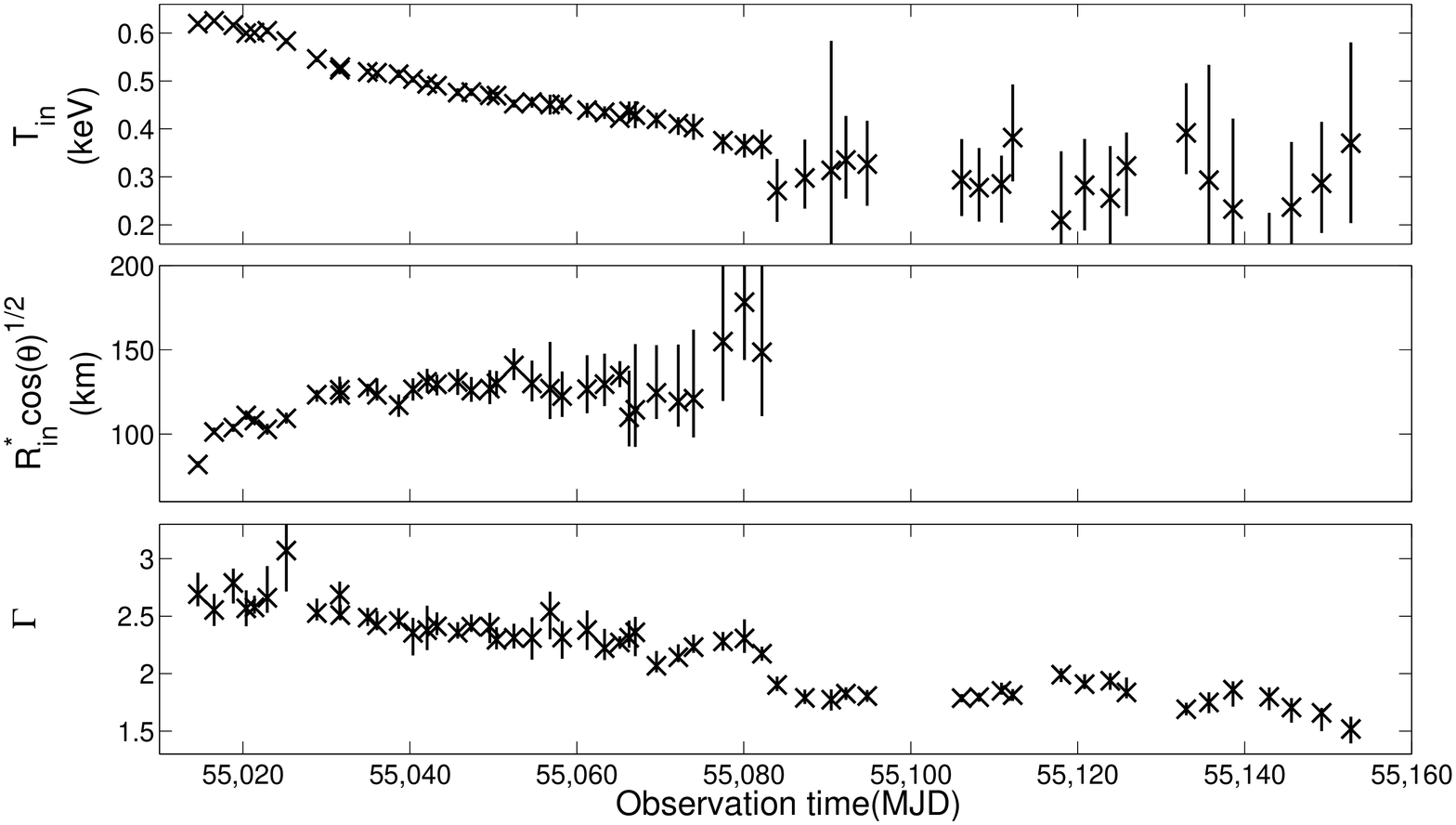}}}
\caption{From top to bottom: The flux of 2.0--20keV energy band, the fractional rms amplitude (0.1 -- 125 Hz),
 the thermal component fraction, the temperature of disc blackbody, the observed inner radius of the disk, and the photon index of power law component. Noticed that only the first 35 observed radius of the disk were plotted because of their lower uncertainties.}
\label{fig:spectra}
\end{center}
\end{figure}

\subsection{Timing analysis}
The power spectrum is a powerful method for probing the timing properties of black holes and other accretion powered X-ray binaries. In order to analyze the timing properties of XTE J1652$-$453, the power spectra were constructed for all the RXTE observations using Ftools/Powspec. The Event-mode data from all available PCUs with 125 $\mu$s time resolution over the nominal 2--60 keV range were used.
First, we extracted the light curves of XTE J1652$-$453 with a time resolution of $\sim$ 4 ms. Then each observation was divided into segments of 32 s, and the Fourier transform of each segment was calculated. Therefore, each power spectrum was generated in the $1/32$--125 Hz frequency range. We used a sum of an constant value, a power law component and several Lorentzian shape if there are QPO like features to fit the power spectra. Then we calculated the integrated fractional rms for each observations in the range of 0.1--125 Hz to quantify their characteristics. An example of the power spectra (obsID 94432--01--08--01) is shown in Figure \ref{fig:powerspec}. The evolution of calculated fractional rms amplitude is present in Figure \ref{fig:spectra}. In addition, we reproduced the power density spectra corrected for poisson noise to investigate the timing properties further if there are any QPO like features. However, no QPOs were found in XTE J1652$-$453.

\setcounter{figure}{4}
\begin{figure}
\begin{center}
\resizebox{1.07\columnwidth}{!}{\rotatebox{-90}{\includegraphics[clip]{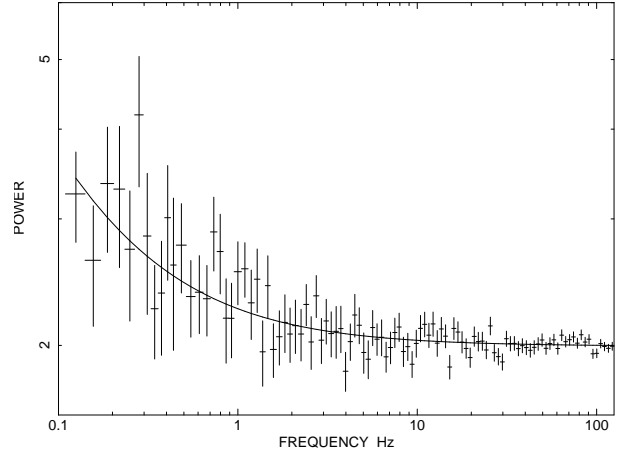}}}
\caption{The power spectrum of XTE J1652$-$453 (obsID 94432--01--08--01). A model consist of an constant value, a power law component was used to fit the spectrum.}
\label{fig:powerspec}
\end{center}
\end{figure}

\subsection{Spectral states and evolution of the source}
The intensity, hardness and spectral parameters evolution histories of XTE J1652$-$453 can be seen in Figure \ref{fig:lightcurve} and Figure \ref{fig:spectra}. Based on the results of the spectral analysis, the evolution of XTE J1652$-$453 can be divided into three distinct parts: the {\sl high/soft}(HS), the {\sl hard intermediate}(HIMS) and the {\sl low/hard}(LH) state. In this work, the {\sl high/soft} state is defined as the set of conditions for which the thermal component fraction is $>85\%$ (2-20keV) with the hardness is  $<0.13$. The {\sl low/hard} state is defined as the power law component contributes $\geq80\%$ (2-20keV) of total flux with the photo index $<2.0$, and the hardness is $>0.75$\citep{McClintock2004, Remillard06}. During the transition from the {\sl high/soft} state to the {\sl low/hard} state, XTE J1652$-$453 experiences the {\sl hard intermediate} state as described in the work of Hiemstra\citep{Hiemstra2010}.

{\sl High/soft} state. It is apparent that during the first seven RXTE observations, the source was in {\sl high/soft} state with a high intensity, low hardness ratio and a comparatively small fraction rms. Only weak power-law noise is observed in the power spectrum. At the beginning of the outburst, the X-ray flux is dominated by the soft component which is believed to be generated from a region of the inner accretion disk. The average temperature of the multicolor disc black body is $\sim 0.6$ keV, and it makes up $\geq85\%$ of 2.0--20keV flux.

{\sl Hard intermediate} state. During the transition form the {\sl high/soft} state to the {\sl low/hard} state, the XTE J1652$-$453 experienced a {\sl hard-intermediate} state. This state is defined to start from the eighth RXTE observation(on 2009 July 16) at which the hardness ratio begins to increase accelerated. It ends on 2009 September 08 (observation 35 in Table 3), when the hardness ratio reaches a basically constant value as well as other spectral parameters such as power law photon index and its contribution to total flux. The intensity and X-ray flux decrease rapidly and the hardness ratio increases significantly during this state.  All the spectral parameters have notable evolutions with time. The proportion of soft component to the total X-ray flux decreases from $\sim85\%$ after the HS state to a few more than $\sim15\%$ at the end of the {\sl hard-intermediate} state. The temperature of multicolor disc blackbody and the photon index of power law component decrease linearly during the HIMS. During this period, the fractional rms has a very clear trend to increase with time as shown in Figure \ref{fig:spectra}.

{\sl Low/hard} state. After the HIMS, XTE J1652$-$453 gets into the LH state. This state is associated with relatively low values of the accretion rate. The intensity and X-ray flux are very low and have a slightly decrease with time. The X-ray flux is dominated by a hard component which makes up more than $80\%$ of total flux. The hardness keeps in an almost constant value, as well as the spectral parameters such as the temperature of disc blackbody and the photon index of power law component.

Overall from the analysis, we can see that during the RXTE observations XTE J1652$-$453 experienced an evolution from the {\sl high/soft} state through the {\sl hard intermediate} state then to the {\sl low/hard} state. It has a clear evolution track from the upper left branch in the HID to the lower right as mentioned in section 3.1. During the transitions, both of the spectral and timing properties have significant changes.

\section{Discussion and summary}
As we know now, many X-ray transients are black hole candidates with similar spectral and timing properties. Basically, these systems are found in two main spectral states: the {\sl high} state which dominated by soft component and the {\sl low} state with a hard spectrum. Additionally, an intermediate state has been recognized with a two-component spectrum, and it can be divided into {\sl hard-intermediate} state and {\sl soft-intermediate} state distinguished in the hardness-intensity diagrams. The X-ray QPOs observed in the power spectra of  black hole candidates in low-mass X-ray binaries are transient phenomena associated with the nonthermal states and state transitions. In particular, it is the sharp changes of the fast variability properties that must be taken as landmarks to separate different states\citep{Belloni05, Belloni09}. Both the low frequency QPOs (0.1--30Hz) and high frequency QPOs (40--450Hz) have been detected in several sources, and many models are developed to describe the nature of QPOs\citep{Zhang2005, Zhang2007, Remillard06}. It is believed that the state transitions are driven by the variable accretion rate\citep{McClintock2004, Remillard06} and some other factors such as the shift in accretion flow geometry\citep{Maccarone2003}. There are multiple models of the spectral states, and many of them invoke the instabilities in the disk\citep{Esin1997, Di Matteo1999, Merloni2001, McClintock2004}.

In this work, we have a systematic study on the spectral and timing properties of XTE J1652$-$453 during its 2009 outburst with RXTE observations. In general, the evolution of spectral parameters shows very similar characteristics to the other black hole systems during the outburst and decay. It has a classic light curve like the 2002 outburst of 4U 1543$-$47\citep{Kalemci05}, and its evolution track on HID is present as the end part of q like diagram\citep{Belloni09, Hiemstra2010}. Based on our analysis, the XTE J1652$-$453 could be divided into three distinct spectral states: the {\sl high/soft} state, the {\sl hard intermediate} state and the {\sl low/hard} state. The mean 2.0--20keV unabsorbed luminosity of XTE J1652$-$453 on 2009 July 04 outburst (MJD $\sim$ 55016) near the peak of the lightcurve is about $6.7\times10^{37}$erg s$^{-1}$ for an assumed distance of 8kpc. During the transition from the {\sl high/soft} state to the {\sl low/hard} state, both the soft flux and the fraction of soft component decrease substantially.

It is very interesting to notice that, a dip appeared in the lightcurve and hardness diagram in Figure \ref{fig:lightcurve} around MJD$\sim55080$ when the source transformed from the {\sl hard intermediate} state to the {\sl hard} state, which is very like the 2002/2003 outburst of GX 339$-$4\citep[see Fig.1 in][]{Belloni05}. At the same time, the soft fraction and photon index of power law component in our spectral analysis had a transient trend to increase. However, the temperature of the multitemperature disc blackbody kept decrease without any significant changes. Therefore, we believe that the transitory increase of soft component fraction and the power law photon index are caused by the decrease of hard X-ray flux. After that, both the soft fraction and power law photon index decreased significantly with a great increase of hardness in a few days. Maybe this dip in the hardness diagram is a mark of the state transition from the {\sl hard intermediate} state to the {\sl hard} state.

In addition, the fractional rms has a clear trend of increase with observation time during XTE J1652$-$453 evolves from the {\sl high/soft} state to the end of {\sl hard-intermediate} state. In the analysis of Hiemstra\citep{Hiemstra2010}, three Lorentzians peaks at different frequencies were present in the power spectrum of the RXTE observation (obsID 94432$-$01$-$04$-$00) which was taken simultaneously with the one of XMM-Newton. However, in our calculation of the RXTE observation with a 4ms time resolution, no significant QPOs were found in the PCA data. In some cases of black hole binaries, the centroid frequencies of QPOs are related to photon energy\citep{Qu2010}. Then we re-calculated the power spectra of the RXTE observation (obsID 94432$-$01$-$04$-$00) in three different energy band (2.06-4.09keV, 4.09-9.81keV and 9.81-20.2keV). However, the QPOs were still not found in these energy band.

At the beginning of the outburst, XTE J1652$-$453 stayed in the {\sl high/soft} state. The derived temperature of disc blackbody and the corresponding observed inner radius R$^{\ast}_{in}$ of the emission area kept an almost constant value during the first 7 observations as shown in Figure\ref{fig:spectra}, so did the soft component fraction. It maybe indicate that the accretion disk stayed in a stable orbit during this period. Assuming that it is the innermost stable circular orbit (ISCO), then the estimated upper limit mass of a Schwarzschild black hole in an assumed distance of 8kpc is $\sim10M_{\odot}$. However, one should bear in mind that the value of the black hole mass derived above depends on the spectral model, the binary system inclination angel and assumed source distance.

During the {\sl hard-intermediate} state of XTE J1652$-$453 2009 outburst, a broad and strong Fe emission line was observed with XMM-Newton with an equivalent width of $\sim450$ eV. The line is consistent with being produced by reflection off the accretion disk, broadened by relativistic effects\citep{Hiemstra2010}. In addition, the diffuse 6.7keV line emission from the Galactic ridge region was approximately derived from the results of GINGA observations of the Galactic Center\citep{Yamauchi1993}. However, no significant Fe lines are observed in this work which maybe due to the poor energy resolution of PCA.

\section{Acknowledgements}
This work was subsidized by the National Natural Science Foundation of China, the CAS key Project KJCX2-YW-T03, and 973 program 2009CB824800, the Natural Science Foundation of China 10733010, 11073021 and 10821061.

\setcounter{table}{2}
\begin{table*}
\begin{minipage}{155mm}
\begin{center}{
\scriptsize
\caption{Spectral parameters of RXTE observations of XTE J1652$-$453, derived using a combined model of the multicolour disc black-body, power-law models with correction for interstellar absorption and a smeared edge. }
\label{tab:line-models}
\begin{tabular}{c c c c c c c c c c c }

\hline\hline
\# &Obs.ID  &Date  &$T_{in}$ &$R^{\ast}_{in}\cos(\theta)^{1/2}$ &$\Gamma$    &$Flux(2.0 -- 20keV)$  &soft fraction  &hardness  &$\chi^{2}$ \\
 &       &MJD   &keV  &km     &   &$10^{-10}$erg s$^{-1}$cm$^{-2}$&& & &\\
\hline
1 &94044-02-01-00  &55014.6  &$0.62^{+0.01}_{-0.01}$ &$82^{+1.9}_{-1.9}$ &$2.69^{+0.19}_{-0.10}$  &53.4  &$88.5\%$  &0.10  &1.00     \\
2 &94044-02-02-00  &55016.6  &$0.63^{+0.002}_{-0.003}$ &$101^{+2.6}_-1.6{}$ &$2.55^{+0.14}_{-0.14}$  &86.7  &$88.0\%$  &0.11  &1.02       \\
3 &94044-02-02-01  &55018.9  &$0.62^{+0.01}_{-0.002}$ &$103^{+2.2}_{-2.4}$ &$2.79^{+0.12}_{-0.18}$  &83.9  &$87.5\%$  &0.10  &1.04        \\
4 &94044-02-02-02  &55020.4  &$0.61^{+0.01}_{-0.01}$ &$111^{+2.9}_{-2.6}$ &$2.57^{+0.16}_{-0.16}$  &81.9  &$87.0\%$  &0.11  &0.97       \\
5 &94044-02-02-03  &55021.4  &$0.60^{+0.01}_{-0.01}$ &$108^{+2.6}_{-2.2}$ &$2.58^{+0.10}_{-0.06}$  &78.9  &$86.3\%$  &0.12  &1.12       \\
6 &94044-02-03-02  &55022.9  &$0.61^{+0.007}_{-0.007}$ &$103^{+3.2}_{-3.1}$ &$2.66^{+0.28}_{-0.13}$  &72.6  &$88.5\%$  &0.10  &0.65       \\
7 &94044-02-03-01  &55025.2  &$0.58^{+0.006}_{-0.006}$ &$109^{+3.4}_{-3.1}$ &$3.07^{+0.30}_{-0.36}$  &66.1  &$88.0\%$  &0.09  &1.26        \\
8 &94044-02-03-00  &55028.9  &$0.55^{+0.006}_{-0.007}$ &$123^{+2.8}_{-4.1}$ &$2.53^{+0.13}_{-0.06}$  &58.3  &$83.5\%$  &0.13  &1.04       \\
9 &94044-02-04-00  &55031.6  &$0.52^{+0.01}_{-0.01}$ &$126^{+7.7}_{-6.7}$ &$2.69^{+0.11}_{-0.15}$  &55.1  &$70.6\%$  &0.20  &0.83       \\
10&94044-02-04-01  &55031.7  &$0.53^{+0.004}_{-0.01}$ &$123^{+7.9}_{-4.8}$ &$2.51^{+0.09}_{-0.04}$  &55.0  &$72.1\%$  &0.21  &0.79       \\
11&94044-02-04-02  &55035.0  &$0.52^{+0.01}_{-0.004}$ &$127^{+2.3}_{-5.0}$ &$2.49^{+0.09}_{-0.07}$  &45.3  &$83.0\%$  &0.14  &1.19        \\
12&94044-02-05-01  &55036.1  &$0.52^{+0.005}_{-0.01}$ &$123^{+9.8}_{-3.4}$ &$2.42^{+0.12}_{-0.04}$  &44.6  &$76.7\%$  &0.19  &1.33        \\
13&94044-02-05-00  &55038.7  &$0.51^{+0.01}_{-0.008}$ &$117^{+6.3}_{-6.8}$ &$2.46^{+0.11}_{-0.07}$  &40.6  &$73.3\%$  &0.21  &1.28        \\
14&94044-02-05-02  &55040.4  &$0.50^{+0.006}_{-0.007}$ &$127^{+6.5}_{-6.6}$ &$2.35^{+0.13}_{-0.20}$  &36.3  &$83.4\%$  &0.15  &0.63        \\
15&94044-02-05-03  &55042.1  &$0.49^{+0.008}_{-0.008}$ &$131^{+7.7}_{-6.4}$ &$2.38^{+0.21}_{-0.17}$  &33.3  &$85.8\%$  &0.13  &0.92        \\
16&94432-01-01-00  &55043.2  &$0.49^{+0.007}_{-0.007}$ &$130^{+5.7}_{-6.5}$ &$2.41^{+0.12}_{-0.08}$  &32.8  &$80.5\%$  &0.17  &1.12        \\
17&94432-01-01-01  &55045.7  &$0.48^{+0.009}_{-0.007}$ &$131^{+7.7}_{-7.5}$ &$2.36^{+0.09}_{-0.05}$  &31.5  &$71.0\%$  &0.24  &1.08        \\
18&94432-01-01-02  &55047.4  &$0.48^{+0.007}_{-0.008}$ &$126^{+8.0}_{-6.7}$ &$2.41^{+0.11}_{-0.04}$  &29.7  &$70.1\%$  &0.24  &1.26        \\
19&94432-01-01-03  &55049.6  &$0.47^{+0.01}_{-0.01}$ &$127^{+11}_{-9.5}$ &$2.41^{+0.12}_{-0.14}$  &26.5  &$72.3\%$  &0.23  &1.26        \\
20&94432-01-02-03  &55050.4  &$0.47^{+0.008}_{-0.007}$ &$130^{+7.4}_{-7.9}$ &$2.29^{+0.11}_{-0.08}$  &24.5  &$80.7\%$  &0.18  &1.28       \\
21&94432-01-02-00  &55052.5  &$0.45^{+0.008}_{-0.008}$ &$141^{+10}_{-8.5}$ &$2.28^{+0.13}_{-0.07}$   &24.2  &$74.5\%$  &0.23  &1.18       \\
22&94432-01-02-01  &55054.6  &$0.46^{+0.01}_{-0.01}$ &$130^{+14}_{-11}$ &$2.31^{+0.18}_{-0.19}$   &21.6  &$74.9\%$  &0.23  &0.89        \\
23&94432-01-02-02  &55056.8  &$0.45^{+0.02}_{-0.02}$ &$127^{+28}_{-18}$ &$2.54^{+0.18}_{-0.24}$   &20.0  &$71.5\%$  &0.24  &0.43    \\
24&94432-01-03-00  &55058.2  &$0.45^{+0.01}_{-0.01}$ &$123^{+14}_{-12}$ &$2.31^{+0.14}_{-0.18}$   &19.4  &$69.3\%$  &0.27  &0.88  \\
25&94432-01-03-01  &55061.2  &$0.44^{+0.01}_{-0.02}$ &$127^{+20}_{-14}$ &$2.38^{+0.17}_{-0.17}$   &17.1  &$69.0\%$  &0.28  &0.69   \\
26&94432-01-03-02  &55063.3  &$0.43^{+0.01}_{-0.01}$ &$130^{+18}_{-13}$ &$2.22^{+0.17}_{-0.10}$   &15.9  &$71.7\%$  &0.27  &0.98   \\
27&94432-01-04-00  &55065.1  &$0.42^{+0.01}_{-0.01}$ &$135^{+8.5}_{-6.9}$ &$2.27^{+0.05}_{-0.05}$  &15.8  &$63.6\%$  &0.33  &0.98  \\
28&94432-01-04-03  &55066.3  &$0.44^{+0.02}_{-0.02}$ &$110^{+27}_{-17}$ &$2.31^{+0.15}_{-0.09}$   &15.4  &$55.7\%$  &0.38  &0.91  \\
29&94432-01-04-01  &55067.0  &$0.43^{+0.03}_{-0.03}$ &$114^{+38}_{-21}$ &$2.36^{+0.13}_{-0.21}$   &14.7  &$54.7\%$  &0.38  &0.59   \\
30&94432-01-04-02  &55069.5  &$0.42^{+0.01}_{-0.02}$ &$125^{+28}_{-15}$ &$2.07^{+0.13}_{-0.06}$   &14.5  &$57.2\%$  &0.41  &0.78   \\
31&94432-01-05-00  &55072.2  &$0.41^{+0.01}_{-0.02}$ &$119^{+34}_{-14}$ &$2.14^{+0.11}_{-0.04}$   &13.9  &$46.0\%$  &0.48  &0.44    \\
32&94432-01-05-01  &55074.0  &$0.40^{+0.03}_{-0.03}$ &$121^{+41}_{-23}$ &$2.23^{+0.11}_{-0.05}$   &13.0  &$42.8\%$  &0.48  &0.93    \\
33&94432-01-05-02  &55077.5  &$0.38^{+0.01}_{-0.03}$ &$155^{+74}_{-35}$ &$2.28^{+0.08}_{-0.08}$   &11.3  &$47.9\%$  &0.46  &0.68    \\
34&94432-01-06-00  &55080.1  &$0.37^{+0.02}_{-0.03}$ &$178^{+81}_{-34}$ &$2.31^{+0.17}_{-0.13}$   &10.0  &$58.5\%$  &0.41  &0.99    \\
35&94432-01-06-01  &55082.2  &$0.37^{+0.03}_{-0.03}$ &$149^{+90}_{-38}$ &$2.17^{+0.06}_{-0.06}$   &10.7  &$39.3\%$  &0.54  &0.75    \\
36&94432-01-06-02  &55084.0  &$0.27^{+0.07}_{-0.06}$ &-- &$1.90^{+0.07}_{-0.05}$   &12.7  &$22.8\%$  &0.78  &0.69    \\
37&94432-01-07-00  &55087.3  &$0.30^{+0.08}_{-0.06}$ &-- &$1.79^{+0.07}_{-0.05}$   &10.7  &$18.3\%$  &0.82  &0.76    \\
38&94432-01-07-01  &55090.5  &$0.31^{+0.27}_{-0.16}$ &-- &$1.77^{+0.09}_{-0.09}$   &9.1  &$15.1\%$  &0.85  &0.86    \\
39&94432-01-08-00  &55092.2  &$0.34^{+0.09}_{-0.08}$ &-- &$1.83^{+0.05}_{-0.05}$   &9.0  &$17.4\%$  &0.81  &0.54    \\
40&94432-01-08-01  &55094.8  &$0.33^{+0.09}_{-0.09}$ &-- &$1.81^{+0.05}_{-0.05}$   &8.4  &$17.0\%$  &0.80  &0.76    \\
41&94432-01-10-00  &55106.1  &$0.29^{+0.08}_{-0.07}$ &-- &$1.79^{+0.04}_{-0.04}$  &9.0  &$14.7\%$  &0.84  &1.20     \\
42&94432-01-10-01  &55108.2  &$0.28^{+0.08}_{-0.07}$ &-- &$1.79^{+0.04}_{-0.04}$   &9.7  &$17.7\%$  &0.82  &0.72    \\
43&94432-01-10-02  &55110.9  &$0.29^{+0.06}_{-0.08}$ &-- &$1.85^{+0.07}_{-0.03}$   &10.1  &$18.0\%$  &0.79  &0.74    \\
44&94432-01-10-03  &55112.2  &$0.38^{+0.11}_{-0.09}$ &-- &$1.81^{+0.05}_{-0.05}$   &9.5  &$12.2\%$  &0.80  &0.84   \\
45&94432-01-09-00  &55118.0  &$0.21^{+0.14}_{-0.12}$ &-- &$1.99^{+0.06}_{-0.06}$   &9.7  &$20.9\%$  &0.76  &0.80    \\
46&94432-01-11-00  &55120.8  &$0.28^{+0.10}_{-0.09}$ &-- &$1.91^{+0.09}_{-0.04}$   &8.7  &$18.1\%$  &0.77  &0.97   \\
47&94432-01-11-01  &55123.9  &$0.26^{+0.11}_{-0.10}$ &-- &$1.94^{+0.07}_{-0.08}$   &7.9  &$17.2\%$  &0.78  &0.99   \\
48&94432-01-11-02  &55125.8  &$0.32^{+0.07}_{-0.10}$ &-- &$1.84^{+0.13}_{-0.05}$   &7.3  &$13.4\%$  &0.80  &1.28   \\
49&94432-01-12-00  &55133.0  &$0.39^{+0.10}_{-0.09}$ &-- &$1.69^{+0.06}_{-0.05}$   &5.7  &$14.5\%$  &0.83  &0.93   \\
50&94432-01-13-00  &55135.7  &$0.29^{+0.24}_{-0.20}$ &-- &$1.75^{+0.09}_{-0.09}$   &5.3  &$13.3\%$  &0.91  &0.64   \\
51&94432-01-13-01  &55138.6  &$0.23^{+0.19}_{-0.13}$ &-- &$1.86^{+0.07}_{-0.15}$   &5.3  &$20.2\%$  &0.85  &0.61   \\
52&94432-01-14-00  &55142.9  &$0.15^{+0.07}_{-0.09}$ &-- &$1.80^{+0.08}_{-0.11}$   &5.3  &$29.2\%$  &0.86  &0.62   \\
53&94432-01-14-01  &55145.6  &$0.24^{+0.14}_{-0.17}$ &-- &$1.70^{+0.08}_{-0.13}$   &4.4  &$14.9\%$  &0.93  &1.04   \\
54&94432-01-15-00  &55149.2  &$0.29^{+0.13}_{-0.10}$ &-- &$1.66^{+0.04}_{-0.16}$   &3.5  &$13.6\%$  &0.93  &0.95   \\
55&94432-01-15-01  &55152.7  &$0.37^{+0.21}_{-0.17}$ &-- &$1.52^{+0.11}_{-0.12}$   &2.9  &$14.4\%$  &0.94  &0.76   \\
\hline \\
\end{tabular}
\normalsize}
\begin{tablenotes}
\item[]Note. -- The OBSID, time, the inner temperature of disk, the observed inner radius of the disk, the power law index, the unabsorbed flux of the 2.0--20keV energy band in the unit of $10^{-10}$erg s$^{-1}$cm$^{-2}$, the soft component fraction, and the hardness are given. The hydrogen column density $N_{H}$ is frozen to be $4.0\times10^{22}$cm$^{-2}$ Here $R^{\ast}_{in}=R_{in}/D_{10}$ and $R_{in}$ is the inner radius of the disk.
\end{tablenotes}
\end{center}
\end{minipage}
\end{table*}

\end{document}